%% file: main.tex
\documentclass[conference]{IEEEtran}
\IEEEoverridecommandlockouts
\usepackage{cite}
\usepackage{amsmath,amssymb,amsfonts}
\usepackage{algorithmic}
\usepackage{graphicx}
\usepackage{textcomp}
\usepackage{xcolor}
\usepackage{todonotes}
\usepackage{comment}
\usepackage[pdftex, colorlinks=true, hyperfootnotes=true, hyperindex=true,
            plainpages=false, pagebackref=false, pdfpagelabels=true, pdfstartview=FitH,
            linkcolor=brown, citecolor=brown, urlcolor=blue,
            bookmarks, bookmarksopen, bookmarksdepth=3]{hyperref}
\def\BibTeX{{\rm B\kern-.05em{\sc i\kern-.025em b}\kern-.08em
    T\kern-.1667em\lower.7ex\hbox{E}\kern-.125emX}}
    
\begin{document}

% \title{Enabling Usage Control in Solid through Blockchain-driven Architecture}
\title{A Blockchain-driven Architecture for\\Usage Control in Solid
\thanks{The work of S.\ Kirrane is funded by the FWF Austrian Science Fund and the Internet Foundation Austria under the FWF Elise Richter and netidee SCIENCE programmes as project number V 759-N. The work of D.\ Basile, C.\  Di~Ciccio and V.\ Goretti was partly supported by projects SERICS (PE00000014) under the NRRP MUR program funded by the EU-NGEU, PINPOINT (B87G22000450001) under the MUR PRIN programme, and by the Sapienza project ``Drones as a Service for First Emergency Response''.}
}

\author{\IEEEauthorblockN{Davide Basile, Claudio Di Ciccio, and Valerio Goretti}
\IEEEauthorblockA{\textit{Department of Computer Science} \\
\textit{Sapienza University of Rome, Italy}% \\ \orcidID{0000-0002-5804-4036}
}
\and
\IEEEauthorblockN{Sabrina Kirrane}
\IEEEauthorblockA{\textit{Department of Information Systems and Operations} \\
\textit{Vienna University of Economics and Business, Austria}% \\\orcidID{0000-0002-6955-7718} 
}
}

\maketitle

\begin{abstract}
Decentralization initiatives like Solid and Digi.me enable data owners to control who has access to their data and to stimulate innovation by creating both application and data markets. %As such, they can be a precious aid for small players in the tech industry market to compete with currently dominant, large organizations. 
Once %Solid 
data owners share their data with others, though, it is no longer possible for them to control how their data are used. To address this issue, we propose a usage control architecture to monitor compliance with usage control policies. To this end, our solution relies on blockchain and trusted execution environments. We demonstrate the potential of the architecture by describing the various workflows needed to realize a motivating use case scenario for data markets. Additionally, we discuss the merits of the approach from privacy, security, integrateability, and affordability perspectives. 
\end{abstract}


%%%%%%%%%%%%%%%%%%%%%%%%%%%%%%%%%%%%%%%%%%%%%
% Introduction 
%%%%%%%%%%%%%%%%%%%%%%%%%%%%%%%%%%%%%%%%%%%%%this

\section{Introduction}\label{sec:intro}
\input{content/introduction}

%%%%%%%%%%%%%%%%%%%%%%%%%%%%%%%%%%%%%%%%%%%%%
% framework
%%%%%%%%%%%%%%%%%%%%%%%%%%%%%%%%%%%%%%%%%%%%%

\section{Motivating Use Case Scenario}\label{sec:motivation}
\input{content/motivation}

%%%%%%%%%%%%%%%%%%%%%%%%%%%%%%%%%%%%%%%%%%%%%
% architecture
%%%%%%%%%%%%%%%%%%%%%%%%%%%%%%%%%%%%%%%%%%%%%

\section{Decentralized Usage Control Architecture}\label{sec:architecture}
\input{content/architecture}

%%%%%%%%%%%%%%%%%%%%%%%%%%%%%%%%%%%%%%%%%%%%%
% instantiation
%%%%%%%%%%%%%%%%%%%%%%%%%%%%%%%%%%%%%%%%%%%%%

\section{An Instantiation of the Architecture}\label{sec:instantiation}
\input{content/instantiation}

%%%%%%%%%%%%%%%%%%%%%%%%%%%%%%%%%%%%%%%%%%%%%
% discussion
%%%%%%%%%%%%%%%%%%%%%%%%%%%%%%%%%%%%%%%%%%%%%

\section{Discussion}\label{sec:discussion}
\input{content/discussion}

%%%%%%%%%%%%%%%%%%%%%%%%%%%%%%%%%%%%%%%%%%%%%
% Conclusion 
%%%%%%%%%%%%%%%%%%%%%%%%%%%%%%%%%%%%%%%%%%%%%

\section{Conclusions}\label{sec:conclusion}
\input{content/conclusion}

%%%%%%%%%%%%%%%%%%%%%%%%%%%%%%%%%%%%%%%%%%%%%
% Biblography
%%%%%%%%%%%%%%%%%%%%%%%%%%%%%%%%%%%%%%%%%%%%%

\bibliographystyle{ieeetr}
\bibliography{library}

\end{document}

%% file: content/introduction.tex
% !TEX root = ../main.tex
%Decentralized initiatives such as Solid~\cite{Solid} and Digi.me~\cite{Digi.me} aim to give data owners more control over their data, while at the same time providing small companies as well as individuals with access to data, which is usually monopolized by centralized platform providers, thus stimulating innovation.
Decentralized projects like \emph{Solid}%~\cite{Solid}
\footnote{\url{https://solidproject.org/}. Accessed: \today.}
and \emph{Digi.me}%~\cite{Digi.me}
\footnote{\url{https://digi.me/}. Accessed: \today.}
seek to increase data owners' control over their data while also giving people and small organizations access to information that is typically managed by centralized platforms.
%Towards this end, the Solid community are developing tools, best practices, and web standards that facilitate ease of data integration and support the development of decentralized social applications based on Linked Data principles.
The Solid community aims to achieve this objective by building web standards and best practices that make data integration simple and encourage the creation of decentralized social apps based on Linked Data concepts. 
%In turn, Digi.me is developing tools and technologies that enable individuals to download their data from centralized platforms such that they can store it in an encrypted personal data store and leverage a variety of applications that can process this data locally on the data owners device. These client side applications are developed by innovative app developers who use the Digi.me software development kit to communicate with the encrypted personal data stores directly.
%
To provide individuals with more control over their data, Digi.me develops technologies thanks to which users can encrypt and collect their information from centralized platforms in personal datastores. % Personal datastores enable people to use a variety of client-side applications that are capable of processing the stored data on their own devices. Developers build these client-side applications using the Digi.me software development kit, which safely enables direct communication with the encrypted personal data stores.

In both cases, there is potential for new data and application markets. 
%However, in order to cater for use case scenarios that involve data sharing, there is a need for tools and technologies that are capable of working with distributed data stores as well as data resources that come with a variety of different terms and conditions specified by data owners.
Protocols that design interactions with distributed data stores are essential to work with various data resources that may come with distinct terms and conditions specified by data owners for data sharing. %and address use cases involving data sharing. 
Those terms and conditions typically come in two forms.
\emph{Access control} takes place \emph{before} granting information access~\cite{akaichi2022usage}. 
\emph{Usage control} extends the former as its enforcement requires \emph{runtime} monitoring of data consumption at a remote location.

%Several works strive to provide more control and transparency with respect to personal data processing by leveraging blockchain distributed application platforms~\cite{DBLP:books/sp/XuWS19}. 
A large body of research work improves control and transparency in personal data processing by utilizing blockchain-based distributed application platforms~\cite{DBLP:books/sp/XuWS19}. Ayoade et al.~\cite{8424682} propose a framework wherein blockchain applications are used to manage access to data that are stored off-chain in a trusted execution environment. %In the proposed framework a blockchain based ledger is used in order to develop an audit trail of data access that provides more transparency with respect to data processing. 
Zhaofeng et al.~\cite{8936349} introduce a secure usage control scheme for Internet of Things (IoT) data that are built upon a blockchain-based trust management approach. Khan et al.~\cite{DBLP:journals/winet/KhanZSAM20} present the \emph{DistU} distributed usage control framework, which applies the UCON$_{ABC}$~\cite{park2004uconabc} model to the Hyperledger Fabric%
\footnote{\url{https://www.hyperledger.org/use/fabric}. Accessed: \today.}
permissioned blockchain. Xiao et al.~\cite{10.1007/978-3-030-59013-0_30} propose a system called \emph{PrivacyGuard},  which leverages blockchain technologies to share usage policies, records resource usage, and monitors policy compliance in a data market scenario. Furthermore, several research studies propose integrations of the Solid protocol with blockchain technologies. 
%Ramachandran et al.~\cite{10.1145/3366424.3385759} demonstrate how together Solid data stores (i.e., pods) and blockchain can be used for trustless verification with confidentiality.
Ramachandran et al.~\cite{10.1145/3366424.3385759} show three possible configurations which combine blockchain with Solid to verify resource integrity, represent resources as smart contracts, and manage crypto wallets through off-chain personal online datastores (\emph{pods}). Cai et al.~\cite{9064776} present a blockchain-assisted system implementing access control policies as a secure authentication mechanism for Solid.
%introduce a secure Solid authentication mechanism, integrating Rivest–Shamir–Adleman (RSA) signatures into permissioned blockchain systems. 
Becker et al.~\cite{DBLP:conf/esws/BeckerVKBK21} propose a blockchain-based payment protocol to build a monetization framework for data stored in Solid personal online datastores. 
Havur et al.~\cite{havur2020greater} show a decentralized layered architecture supporting the intersection of the SPECIAL%
\footnote{\url{https://ai.wu.ac.at/policies/policylanguage/} Accessed: \today.}
policy language with Solid standards integrated into personal online datastores.
%Whereas, Havur et al.~\cite{havur2020greater} show a layered decentralised architecture discuss the level of control, transparency and compliance provided.

Despite these efforts, Solid currently only supports basic access control, and thus it is not possible to ensure that data consumers adhere to usage restrictions specified by data owners. 

To overcome this limitation, we propose a decentralized usage control architecture that resorts to a blend of blockchain applications and trusted execution environments. We extend the state of the art by demonstrating (i) how \emph{blockchain oracles}~\cite{basile2021enhancing} allow for seamless communication between these entities, and (ii) how Solid applications~\cite{sambra2016solid} can be enhanced with usage control mechanisms. In the proposed architecture, users' data are kept in Solid personal online datastores. Access is administered through a component named \emph{pod manager}. The usage control is handled by blockchain executable applications that are capable of (i) recording where data resides, (ii) declaring what the usage restrictions are, and (iii) monitoring compliance with these policies. Applications that leverage data stored in Solid pods run in a \emph{trusted execution environment}~\cite{DBLP:conf/trustcom/SabtAB15}, which enables users to revoke access if data consumers do not adhere to the %terms and conditions specified by data owners in the form of 
usage policies. Finally, blockchain oracles enable pod managers and trusted execution environments to communicate with the blockchain and vice versa. We illustrate the application of our architecture and highlight its effectiveness in the %the effectiveness of the proposed architecture is demonstrated %with the help of a data market motivating use case scenario. 
% through its application %on a motivating use case scenario 
in the context of data markets.

Next, Section~\ref{sec:motivation} describes a motivating scenario we employ as a running example throughout this paper. Section~\ref{sec:architecture} presents the software architecture at the core of our solution. Section~\ref{sec:instantiation} illustrates the application of our approach to the motivating scenario. Section~\ref{sec:discussion} evaluates our approach through the lens of four key properties. Finally, Section~\ref{sec:conclusion} concludes the paper and outlines possible endeavors for future work.

%% file: content/motivation.tex
% !TEX root = ../main.tex
%DistExchange
% Data trading across decentralized datastores is now made easier by a new decentralized data market. % called \emph{DistExchange}. 
Alice and Bob sign up for a new decentralized data market service for data trading across datastores. Their accounts include contact details, subscription details, and a username and password for the service. %Bob already has a personal datastore that contains various pieces of information. However, he wants to make part of this data accessible via the market. 
They set up a personal datastore on a server of their choosing, wherein they add the data that they would like to trade. % into her datastore. Both
Alice and Bob employ usage policies to set usage restrictions on their data. Bob's dataset contains medical data to be used only for medical purposes. Alice's dataset contains internet-browsing datasets, which must be deleted one month after their storage. Alice and Bob send metadata %with respect to the
associated with the data that they would like to trade alongside the usage policies to the decentralized data market. 

Alice is a researcher in the healthcare domain. She is interested in Bob's medical dataset. %, which may only be used for medical purposes. 
She asks the service for a data reference and a certificate proving she has paid the market fee. Alice uses the reference to contact Bob's personal datastore and check the certificate's validity. Thereafter, it returns Bob's medical dataset and the associated usage policy. Similarly, Bob, a web data analyst, wishes to retrieve Alice's internet-browsing dataset from her personal online datastore. Alice and Bob only use the data obtained from the market on their trusted devices (which is part of the terms and conditions stipulated by the market), in which a trusted software component enforces policies. This ensures that Alice's dataset is deleted after one week of usage and Bob's healthcare data are only used for medical purposes. Alice asks the market service to check that the usage policy associated with her datasets is being adhered to. In this case, trusted devices storing her resources provide her with evidence of the policies' compliance. At any point, Alice and Bob can change the rules associated with their datasets. In particular, after two days, Alice changes the maximum storage time of her internet-browsing data to one week. In the meantime, Bob modifies the allowed purpose of use of his medical resources to academic pursuits. Trusted devices guarantee ongoing policies update after the information retrieval, thus catching Alice and Bob's policy updates from the market service. As a result, Alice's data are erased from Bob's device after the new expiry time lapses. As Alice is using an application in the medical research domain for a university hospital, changes do not affect her access grants.
% \todo[inline]{Scenarios are not about what actors can do but about what they do. To be changed. Again.}

%\todo[inline]{As is, this is not a motivating scenario. Missing: why TEE at all if neither Alice nor Bob change usage policies or see any breach in the usage policies? If only medical data are exchanged, then why not use dedicated applications rather than a general purpose one?}

%% file: content/architecture.tex
% !TEX root = ../main.tex
%\begin{comment}
\begin{figure*}[!t]
\includegraphics[width=\textwidth]{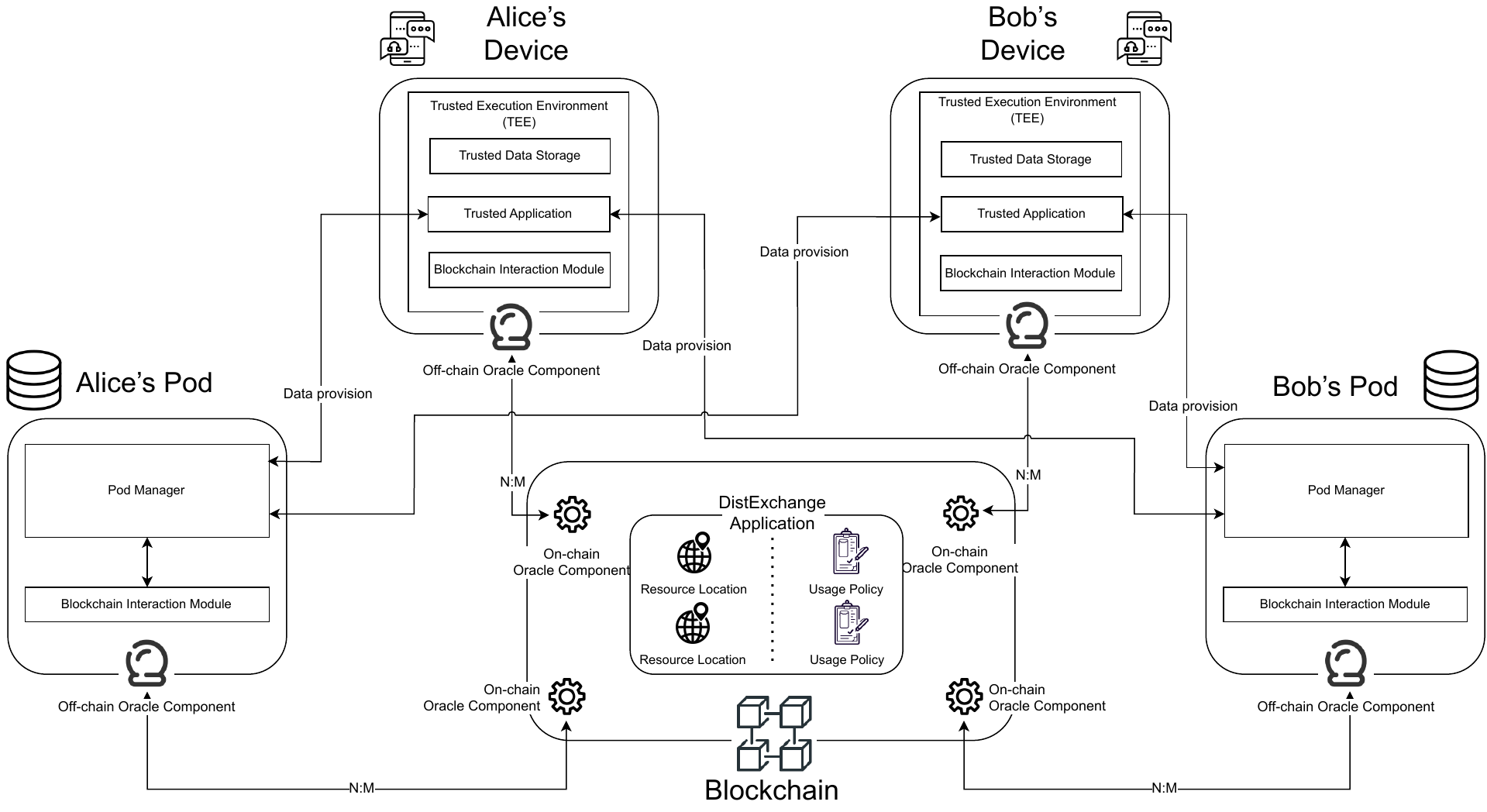}
\caption{A Decentralised Usage Control Architecture}
\label{fig:architecture}
\end{figure*}
%\end{comment}

\begin{comment}
Notes about oracles:
\textbf{push-in:} used to insert and update policies (or change existing ones); instantiation/deployment of a contract instance is a part of it;
\textbf{pull-out:} used to verify / read the policy;
\textbf{push-out:} erases or alters the content of cache upon a change in the policy (say, Alice is not granted any longer access to Bob's photo $\to$ out of Alice's cache!);
\textbf{pull-in:} used to retrieve the data/policy associated to IDs that do not represent directly the real objects or users (say, an account requests the grants to access a given resource: who is behind this account?).
\end{comment}

%Decentralised usage control architecture.
To cater for our motivating use case scenario, we propose an architecture that extends Solid with usage control capabilities. More specifically, we build upon the existing Solid infrastructure, which we enhance to (i) continuously monitor compliance with usage policies and (ii) enforce the fulfillment of usage policy obligations after access to data has been granted. %The various components of the proposed architecture, which is depicted in 
Figure~\ref{fig:architecture} depicts the proposed architecture. We describe it in detail below.

\subsection{Pods, Pod Managers, and the Solid Protocol} 
Our architecture extends the Solid protocol, whose main goal is to support decentralized data storage and application development~\cite{sambra2016solid}. Solid applications communicate with personal data stores called \texttt{Pod}s, according to the Solid communication rules, via \texttt{Pod Manager}s. The \texttt{Pod Manager} is %essentially 
a web application %that is capable of processing HTTP requests 
that allow %authenticated and/or unauthenticated 
users to retrieve, modify and control data that are stored in a Solid \texttt{Pod}. Thus, the \texttt{Pod Manager} determines whether %if 
access %should 
can be granted %or not via 
by checking the access control policies that are stored locally. % that specify if the entity that issued the request is permitted to perform the requested operation. 

However, once data are retrieved from the Solid \texttt{Pod}, %using a read operation, 
it is not possible for Solid to control how data are subsequently used. Thus, we combine the Solid infrastructure with a distributed %\texttt{Blockchain} 
blockchain application that facilitates usage control after data have been retrieved. 

\subsection{Blockchain, DistExchange Application, and Usage Policies}
Modern
%\texttt{Blockchain}s 
blockchain technologies offer trusted and secure environments not only for classical data storage but also for the execution of applications that run on distributed virtual machines~\cite{cai2018decentralized}. The correctness of the executed code is validated by the consensus mechanism of the %\texttt{Blockchain} itself. 
blockchain. 

In Fig.~\ref{fig:architecture}, we enclose the multiple software elements we deploy on the blockchain infrastructure in a dedicated macro-component labeled as \texttt{Blockchain}, which we leverage for multiple aims. First of all, we resort to its ledger to store references to the physical location of Solid \texttt{Pod}s, as well as specific \texttt{Resource Location} and applicable \texttt{Usage Policies}. Additionally, we resort to the distributed virtual machine running \emph{smart contracts} to develop a \texttt{DistExchange Application} (\texttt{DE App}) that is capable of monitoring compliance with usage control policies. For instance, a \texttt{Usage Policy} may specify temporal obligations that state the duration of usage for a particular resource (e.g., the one-week expiry of Alice's web data) and purpose obligations that constrain resource usage to a given purpose (e.g., medical research as the sole access aim for Bob's data). %, social applications, etc.). 

The \texttt{DE App} is responsible for monitoring compliance with every \texttt{Usage Policy} and detecting policy violations. It relies on the \texttt{Trusted Execution Environment} hosted by data consumer devices to enforce \texttt{Usage Policy}.

\subsection{Trusted Execution Environment} 
A \texttt{Trusted Execution Environment} is composed of hardware and software that ensures the protection of sensitive data by providing isolated execution, application integrity, and data confidentiality~\cite{DBLP:conf/trustcom/SabtAB15}. 
A \texttt{Trusted Application} is a software object running in a \texttt{Trusted Execution Environment}. Our infrastructure imposes that Solid client requests are generated by \texttt{Trusted Application}s. A copy of the requested data is stored locally and managed by the \texttt{Trusted Execution Environment} through the \texttt{Trusted Data Storage}. %\texttt{Trusted application}s are the only way to use resources that are stored on data consumer devices. 
Local access to the \texttt{Trusted Data Storage} is controlled by the \texttt{Trusted Execution Environment} according to the \texttt{Usage Policy}. For instance, %suppose the \texttt{Usage Policy} of a retrieved resource contains a temporal obligation, which specifies that the data must be deleted after one month of usage.
consider the temporal obligation on Alice's data.
In this case, the \texttt{Trusted Execution Environment} %understands the condition specified in the policy and 
automatically deletes the resource from the \texttt{Trusted Data Storage} after %the specified period of time. 
one week has passed, as per the policy. 
%
%Another important aspect that is managed by the 
The \texttt{Trusted Execution Environment} logs resource usage, too. %is resource usage logging. 
This feature facilitates policy monitoring whereby the \texttt{Blockchain} regularly interacts with the \texttt{Trusted Execution Environment} in order to ensure that usage policies are being adhered to.
For instance, Bob can routinely check who the granted users to his data are and what use they are making of his shared information.
\texttt{Pod Manager}s and \texttt{Trusted Execution Environment}s communicate with the \texttt{Blockchain} and vice versa via blockchain oracles.

\subsection{Communication via Blockchain Oracles}
Given that blockchains are closed environments, applications running in the blockchain ecosystem cannot natively communicate with entities located outside the network. For this reason, communication mechanisms called oracles are needed in order to connect the on-chain to the off-chain world~\cite{DBLP:conf/bpm/MuhlbergerBFCWW20}. Oracles are trusted entities used to facilitate data flow from the on-chain apps to real-world software and vice versa. We classify oracles according to two criteria: flow direction (in-bound/out-bound) and data operation (pull-based/push-based). Considering these criteria, it is possible to distinguish four types of oracles: \emph{push-in}, \emph{push-out}, \emph{pull-in}, and \emph{pull-out}. To be realized, oracles are split into two core parts. One lies off-chain and the other lies on-chain~\cite{basile2021enhancing}. 

In the proposed architecture, the off-chain entities that communicate with the \texttt{Blockchain} are \texttt{Pod Managers} and the \texttt{Trusted Execution Environment}s hosted on data consumer devices. These applications interact with the \texttt{Blockchain} via \texttt{Blockchain Interaction Module}s and the respective \texttt{Off-chain Oracle Component}s. We assume that each off-chain entity has the credentials necessary to sign transactions and send data to the \texttt{Blockchain}. % Similarly, the \texttt{DE App} receives or sends data from and to the \texttt{On-chain Oracle Component} by implementing interfaces.

%% file: content/instantiation.tex
%\begin{comment}
\begin{figure*}[!t]
\includegraphics[width=\textwidth]{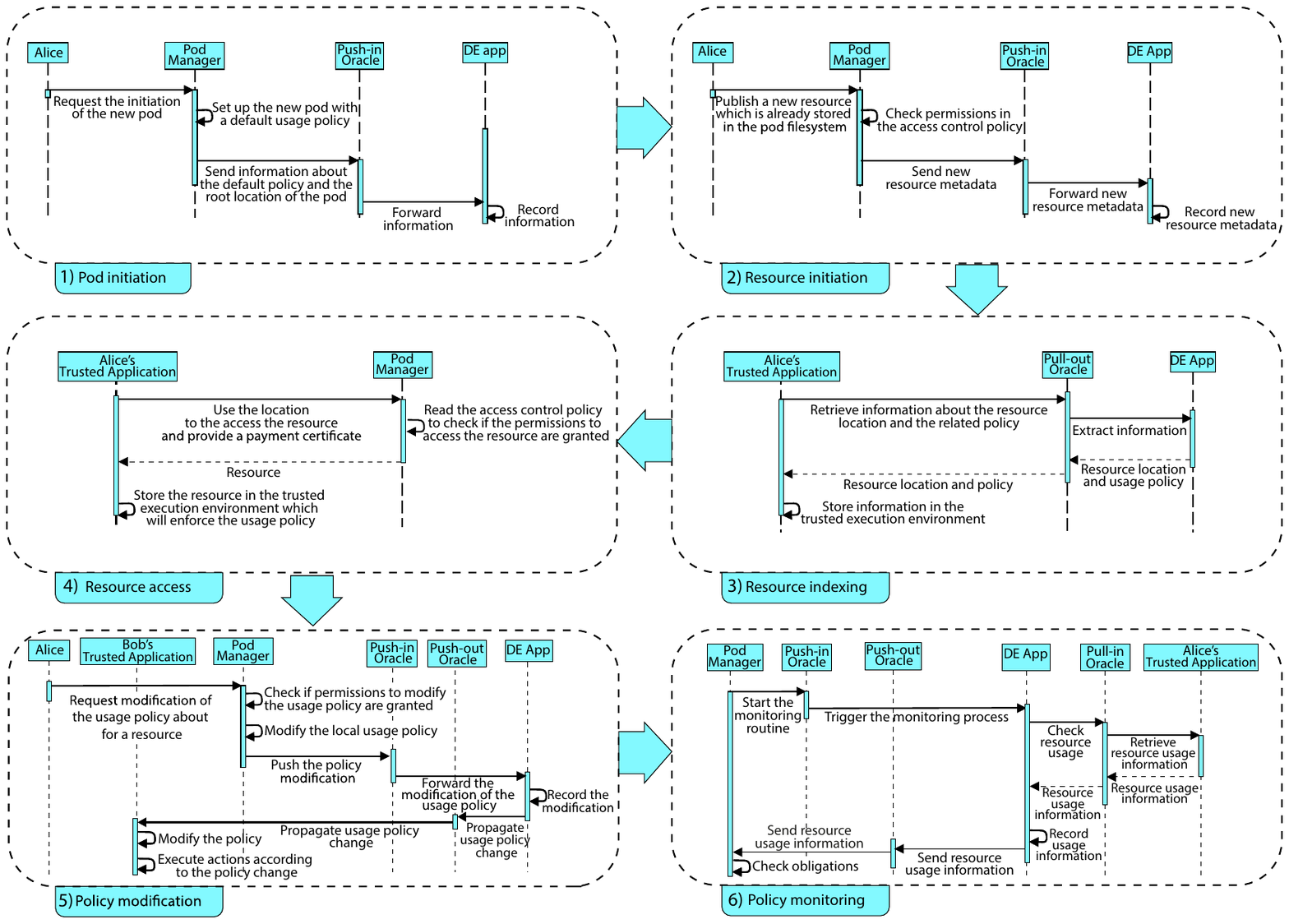}
\caption{Decentralised Usage Control Architecture Processes.}
\label{fig:instantiation}
\end{figure*}
%\end{comment}

We demonstrate the effectiveness of the proposed architecture by revisiting our motivating use case scenario. To this end, we analyze the data flows and interactions among components, which we separate into subsequent processes. %, with a special as well as how they make use of blockchain oracles. 
We specifically focus on the interaction between \texttt{Pod Manager}s, \texttt{Trusted Execution Environment}s and the \texttt{DE App}. Thus, we do not go into specifics in terms of setting up the market and catering for new registrations. The various processes, which are depicted in Fig.~\ref{fig:instantiation}, are described in detail below. 

\subsubsection{\textbf{Pod initiation}} 
Once Alice and Bob have registered with the \texttt{DE App}, they need to link their Solid \texttt{Pod}s to their respective data market accounts. 
The process starts when \texttt{Alice} makes a request to the \texttt{Pod Manager} to initialize a new \texttt{Pod}. The \texttt{Pod Manager} sets up the \texttt{Pod} with its default policy (e.g., only subscribed users have access to the data). Thereupon, it invokes a \texttt{Push-in Oracle} to send information about the \texttt{Pod}'s web reference and its default policy to the blockchain smart contract that is part of the \texttt{DE App}. The process for Bob is analogous.  

\subsubsection{\textbf{Resource initiation}}
The resource initiation process is used in order to add a new resource to the \texttt{DE App}. The process begins when \texttt{Alice} asks her \texttt{Pod Manager} to add a resource to the market that has already been uploaded in her \texttt{Pod}'s filesystem via the Solid protocol. The \texttt{Pod Manager} first checks that \texttt{Alice} is permitted to perform this action. If so, the \texttt{Pod Manager} uses the \texttt{Push-in Oracle} to forward the necessary metadata to the \texttt{DE App} (i.e., a reference to the resource and possibly a resource-specific \texttt{usage policy}), which adds the resource's metadata to the index and publishes the applicable \texttt{Usage Policy}.

\subsubsection{\textbf{Resource indexing}}
With this process, users retrieve a link to a resource that is initialized in the \texttt{DE App}. Alice is interested in the medical data that Bob has added to the market. Given that Alice does not know the exact web location of the resource, she asks the \texttt{DE App} for a link to it alongside the corresponding \texttt{usage policy}. 
The process is initiated when Alice requests information about the resource (i.e., the aforementioned web link and policy). Alice's \texttt{Trusted Application} generates the request, running in the \texttt{Trusted Execution Environment}. It uses the \texttt{Pull-out Oracle} to read this piece of information directly from the \texttt{DE App} running in the \texttt{Blockchain}. The retrieved information is stored in Alice's \texttt{Trusted Execution Environment} and can subsequently be used to retrieve the resource physically.

\subsubsection{\textbf{Resource access}}
The resource access process allows data consumers to retrieve information stored in a Solid \texttt{Pod}. In order to collect Bob's data, Alice's \texttt{Trusted Application} makes a request from within the \texttt{Trusted Execution Environment} to the \texttt{Pod Manager}. The request includes a certificate that proves she has paid the market fee. The \texttt{Trusted Application} provides the \texttt{Pod Manager} with a reference for the resource that it obtained from the \texttt{DE App} via the above process of resource indexing. The \texttt{Pod Manager} first checks that Alice is permitted to perform the read action. If so, the \texttt{Pod Manager} returns the resource to Alice's \texttt{Trusted Application}. In turn,  Alice's \texttt{Trusted Application} stores it within its \texttt{Trusted Data Storage}.

\subsubsection{\textbf{Policy modification}}
The policy modification process enables users to update usage policies after resources have been deployed to the \texttt{DE App}. For instance, \texttt{Alice} shortens the time lapse for the usage of her internet browsing data to one week, whereas it was initially set to one month. Here we assume that such updates are permitted according to the general rules of the market. \texttt{Alice} makes a request to her \texttt{Pod Manager} to change the \texttt{Usage Policy} for that resource. The \texttt{Pod Manager} checks whether \texttt{Alice} is granted the permission to change the policy. If so, it proceeds with the update locally. The \texttt{Pod Manager}, then, uses the \texttt{Push-in Oracle} to send the updated policy to the \texttt{DE App}, which replaces the policy accordingly. The \texttt{DE App} uses a \texttt{Push-out Oracle} to notify those users that have a copy of the resource (e.g., Bob) that the policy has been updated. The \texttt{Trusted Application}s bearing a copy of the resource (e.g., Bob's \texttt{Trusted Application}) update their local policies, check if the change requires any actions to be executed locally, and if so, execute them. In Bob's case, the consequent action to be taken is the erasure of the collected data if the check happens after one week from the first download. Notice that this mechanism is automatically enforceable since data are entirely and solely stored in the \texttt{Trusted Data Storage} as described above.

\subsubsection{\textbf{Policy monitoring}}
The policy monitoring process regularly checks usage policy compliance once data are accessed.
The \texttt{Pod Manager} uses the \texttt{Push-in Oracle} to start the monitoring (for instance, via a scheduled job). The \texttt{Push-in Oracle} forwards the request to the \texttt{DE App}, which in turn communicates with all devices that have a copy of the resource in their \texttt{Trusted Execution Environment} via the \texttt{Pull-in Oracle}. The \texttt{Pull-in Oracle}, then, requests evidence that the usage policies are being adhered to. The \texttt{Push-out Oracle} is subsequently required by the \texttt{DE App} to send the pieces of evidence gathered from the various trusted applications (for instance, Alice's \texttt{Trusted Application}) to the \texttt{Pod Manager} that initiated the policy monitoring process. 

%% file: content/discussion.tex
% !TEX root = ../main.tex

%In this section, we broaden our discussion on the effectiveness of the proposed decentralised usage control architecture with a particular focus on: privacy, security, integrateability, and affordability.
In the following, we expand the discussion of our decentralized usage control architecture, paying special attention to the properties of privacy, security, integrateability and affordability.

\subsubsection{\textbf{Privacy}} 
According to the Solid protocol, data owners decide which entities (authenticated or unauthenticated) can access their resources via Access Control Lists (ACLs). This requirement has a significant impact on privacy and data confidentiality, as the need to subscribe to terms and conditions specified by applications is eliminated. However, Solid principles entail that data are kept in specific user-trusted datastores. Such a design choice can represent a limitation for computationally intensive web applications, which are forced to retrieve data from several data sources. 

The establishment of usage control through \texttt{Blockchain applications} and \texttt{Trusted Execution Environment}s allows data owners to keep control over their data (even after data consumers have obtained copies thereof) and further supports the Solid principle of data ownership. At the same time, \texttt{Trusted Execution Environment}s facilitate compliant data storage and resource usage by implementing usage policy enforcement and related obligations. After the resource retrieval, \texttt{Trusted Application}s benefit from locally stored data (as long as the \texttt{Usage Policy} permit it) without the need to constantly communicate with Solid \texttt{Pod}s, which leads to significant improvements in latency and scalability.

The most critical issue regarding confidentiality relates to the blockchain metadata, which are publicly exposed in most cases. Public blockchains offer public ledgers that are fully readable by every node of the network. In our setting, this availability implies that all users can read usage policies and resource locations. Although making this information public can be desirable on occasions,  %In some use cases it may be desirable to have this data public, there may also be a need to encrypt data stored in the blockchain such that only authorized parties (those that have access to the decryption key) can read this metadata~\cite{pan2011survey}.
data owners might request that only authorized parties (e.g., those with access to the decryption key) can access this information. Typically, approaches that achieve this objective in a blockchain context are based on encryption~\cite{pan2011survey,DBLP:conf/bpm/MarangoneCW22}. 

\subsubsection{\textbf{Security}}
In a decentralized web environment, the lack of a central authority increases the chances that malicious users make unauthorized use of data and metadata managed by the infrastructure.
However, the integrity of user data residing in \texttt{Pod}s is already guaranteed by the Solid protocol through access control policies. %The metadata stored in the blockchain (usage policies and resource locations) are protected from unauthorized updates through the consensus mechanism of the blockchain platform and its distribute nature, which makes the stored data immutable.
The blockchain's consensus algorithm and its distributed nature protect the stored metadata (resource locations and usage policies) from unauthorized modifications, making this information tamper-proof. Moreover, methods through which the state of smart contracts is changed can be invoked only by signing transactions with auditable digital signatures. The \texttt{Trusted Execution Environment} provides a separate environment for code execution and data storage. 
%It has already shown its effectiveness in terms of preventing the injection of malicious code coming from the operating system of the client's machine~\cite{DBLP:conf/trustcom/SabtAB15}, which could jeopardize the integrity of the stored resources or the local representation of usage policies. 
Studies such as the one conducted by Sabt et al.~\cite{DBLP:conf/trustcom/SabtAB15} already showed the effectiveness of these technologies in preventing the execution of malicious code from the operating system's machine, which could compromise the integrity of resources and usage policies stored inside the \texttt{Trusted Execution Environment}. Interactions between the various components that could lead to the modification of resources or usage policies are managed via off-chain and on-chain oracle components, which are able to enact secure information exchange between the blockchain and outer parties~\cite{trustworthyOracles}.

The availability of the \texttt{DE app} is preserved by the distributed nature of the blockchain. % which runs on top of multiple nodes. 
If an attack succeeds in bringing down one of the nodes, the blockchain ecosystem can continue to operate by relying on the rest of the nodes. However, both the Solid \texttt{Pod}s and the \texttt{Trusted Execution Environment}s hosted on user devices need to adopt best practices in terms of hardware and software security to guarantee communication with the blockchain platform.

\subsubsection{\textbf{Integrateability}}
One of the requirements that steer our design process is the need to easily integrate our architecture with the existing Solid ecosystem, so that pod data management functionality could be extended to cater to usage control. \texttt{Pod}s interact with blockchain applications via a plug-in module, which enables subscription, usage policy specification, and resource indexing. On the client side, an additional requirement on the hardware is set by the fact that \texttt{Trusted Execution Environment}s rely on separation kernel methodologies through hardware support~\cite{DBLP:conf/trustcom/SabtAB15}. However, this kind of technology is supported via various extensions to existing operating systems.

\subsubsection{\textbf{Affordability}}
Public blockchains use the tamper-proof register feature to define cryptocurrencies whose transactions are stored in blocks. The execution of on-chain code requires that cryptocurrencies are spent, depending on the computational effort required by the run of the code. 

Resorting to a public blockchain, users of our infrastructure would make a payment to interact with the blockchain metadata through transactions. The market scenario can justify the costs involved in our architecture. A subscription-based business model could offer an incentive mechanism that allows users to overcome the sharing costs and earn a remuneration upon access to their data. Therefore, blockchain applications provide an easy way to guarantee a market profit redistribution to users, proportionately to the accesses granted to their data. The economic incentives are out of scope for this paper and pave the path for future work.

%% file: content/conclusion.tex
% !TEX root = ../main.tex

Motivated by the need to ensure that data consumers adhere to usage restrictions specified by Solid data owners, we proposed a decentralized usage control web architecture that extends existing Solid access control mechanisms to cater to usage control. The effectiveness of the proposed architecture is demonstrated with the help of a motivating use case scenario in the context of data markets. Additionally, we examined the proposed architecture from privacy, security, integrateability, and affordability perspectives. 

Future work includes the integration of a policy language that can be used to specific usage policies at different levels of granularity. We are also interested in the study and design of economic mechanisms supporting the data market adoption. %Therefore, t
The proposed architecture generalizes the blockchain concept, although a wide variety of technologies are currently available. Following the comparative methodology proposed in \cite{Basile.etal/JIPE2023:BCEI}, we plan to instantiate a specific blockchain technology that meets the technological requirements evidenced by our decentralized usage control scenario~\cite{Basile.etal/FBloc2023:BlockchainResourceGovernanceDecentralizedWeb}. An analogous analysis will be applied to the multitude of trusted execution environment technologies available, including Intel SGX~\cite{costan2016intel}, TrustZone~\cite{pinto2019demystifying} and OpenTEE~\cite{mcgillion2015open}. The instantiation process will allow us to evaluate the architecture from the perspectives of performance, scalability, and robustness.

%% file: main.bbl
\begin{thebibliography}{10}

\bibitem{akaichi2022usage}
I.~Akaichi and S.~Kirrane, ``Usage control specification, enforcement, and
  robustness: A survey,'' {\em arXiv preprint arXiv:2203.04800}, 2022.

\bibitem{DBLP:books/sp/XuWS19}
X.~Xu, I.~Weber, and M.~Staples, {\em Architecture for Blockchain
  Applications}.
\newblock 2019.

\bibitem{8424682}
G.~Ayoade, V.~Karande, L.~Khan, and K.~Hamlen, ``Decentralized {IoT} data
  management using blockchain and trusted execution environment,'' in {\em
  IRI}, pp.~15--22, 2018.

\bibitem{8936349}
M.~Zhaofeng, W.~Lingyun, W.~Xiaochang, W.~Zhen, and Z.~Weizhe,
  ``Blockchain-enabled decentralized trust management and secure usage control
  of iot big data,'' {\em IEEE Internet Things}, vol.~7, no.~5, pp.~4000--4015,
  2020.

\bibitem{DBLP:journals/winet/KhanZSAM20}
M.~Y. Khan, M.~F. Zuhairi, T.~A. Syed, T.~G. Alghamdi, and J.~A.
  Marmolejo{-}Saucedo, ``An extended access control model for permissioned
  blockchain frameworks,'' {\em Wirel. Networks}, vol.~26, no.~7,
  pp.~4943--4954, 2020.

\bibitem{park2004uconabc}
J.~Park and R.~Sandhu, ``The uconabc usage control model,'' {\em ACM
  transactions on information and system security (TISSEC)}, vol.~7, no.~1,
  pp.~128--174, 2004.

\bibitem{10.1007/978-3-030-59013-0_30}
Y.~Xiao, N.~Zhang, J.~Li, W.~Lou, and Y.~T. Hou, ``Privacyguard: Enforcing
  private data usage control with blockchain and attested off-chain contract
  execution,'' in {\em Computer Security -- ESORICS 2020} (L.~Chen, N.~Li,
  K.~Liang, and S.~Schneider, eds.), pp.~610--629, 2020.

\bibitem{10.1145/3366424.3385759}
M.~Ramachandran, N.~Chowdhury, A.~Third, J.~Domingue, K.~Quick, and M.~Bachler,
  ``Towards complete decentralised verification of data with confidentiality:
  Different ways to connect solid pods and blockchain,'' in {\em Companion
  Proceedings of the Web Conference 2020}, p.~645–649, 2020.

\bibitem{9064776}
T.~Cai, Z.~Yang, W.~Chen, Z.~Zheng, and Y.~Yu, ``A blockchain-assisted trust
  access authentication system for solid,'' {\em IEEE Access}, 2020.

\bibitem{DBLP:conf/esws/BeckerVKBK21}
H.~Becker, H.~Vu, A.~Katzenbach, C.~H. Braun, and T.~K{\"{a}}fer, ``Monetising
  resources on a solid pod using blockchain transactions,'' in {\em The
  Semantic Web: {ESWC} 2021 Satellite Events}, pp.~49--53, 2021.

\bibitem{havur2020greater}
G.~Havur, M.~Vander~Sande, and S.~Kirrane, ``Greater control and transparency
  in personal data processing,'' in {\em ICISSP}, 2020.

\bibitem{basile2021enhancing}
D.~Basile, V.~Goretti, C.~Di~Ciccio, and S.~Kirrane, ``Enhancing
  blockchain-based processes with decentralized oracles,'' in {\em {BPM}
  (Blockchain and {RPA} Forum)}, pp.~102--118, 2021.

\bibitem{sambra2016solid}
A.~V. Sambra, E.~Mansour, S.~Hawke, M.~Zereba, N.~Greco, A.~Ghanem,
  D.~Zagidulin, A.~Aboulnaga, and T.~Berners-Lee, ``Solid: a platform for
  decentralized social applications based on linked data,'' 2016.

\bibitem{DBLP:conf/trustcom/SabtAB15}
M.~Sabt, M.~Achemlal, and A.~Bouabdallah, ``Trusted execution environment: What
  it is, and what it is not,'' in {\em TrustCom/BigDataSE/ISPA}, pp.~57--64,
  2015.

\bibitem{cai2018decentralized}
W.~Cai, Z.~Wang, J.~B. Ernst, Z.~Hong, C.~Feng, and V.~C. Leung,
  ``Decentralized applications: The blockchain-empowered software system,''
  {\em IEEE Access}, vol.~6, pp.~53019--53033, 2018.

\bibitem{DBLP:conf/bpm/MuhlbergerBFCWW20}
R.~M{\"{u}}hlberger, S.~Bachhofner, E.~C. Ferrer, C.~{Di Ciccio}, I.~Weber,
  M.~W{\"{o}}hrer, and U.~Zdun, ``Foundational oracle patterns: Connecting
  blockchain to the off-chain world,'' in {\em {BPM} (Blockchain and {RPA}
  Forum)}, pp.~35--51, 2020.

\bibitem{pan2011survey}
J.~Pan, S.~Paul, and R.~Jain, ``A survey of the research on future internet
  architectures,'' {\em IEEE Commun. Mag.}, vol.~49, no.~7, pp.~26--36, 2011.

\bibitem{DBLP:conf/bpm/MarangoneCW22}
E.~Marangone, C.~{Di Ciccio}, and I.~Weber, ``Fine-grained data access control
  for collaborative process execution on blockchain,'' in {\em {BPM}
  (Blockchain and {RPA} Forum)}, pp.~51--67, 2022.

\bibitem{trustworthyOracles}
H.~Al-Breiki, M.~H.~U. Rehman, K.~Salah, and D.~Svetinovic, ``Trustworthy
  blockchain oracles: Review, comparison, and open research challenges,'' {\em
  IEEE Access}, vol.~8, pp.~85675--85685, 2020.

\bibitem{Basile.etal/JIPE2023:BCEI}
D.~Basile, I.~D’Adamo, V.~Goretti, and P.~Rosa, ``Digitalizing circular
  economy through blockchains: The blockchain circular economy index,'' {\em J.
  Ind. Prod. Eng.}, vol.~40, no.~4, pp.~233--245, 2023.

\bibitem{Basile.etal/FBloc2023:BlockchainResourceGovernanceDecentralizedWeb}
D.~Basile, C.~Di~Ciccio, V.~Goretti, and S.~Kirrane, ``Blockchain based
  resource governance for decentralized web environments,'' {\em Frontiers in
  Blockchain}, vol.~6, p.~1141909, May 2023.

\bibitem{costan2016intel}
V.~Costan and S.~Devadas, ``Intel {SGX} explained,'' {\em Cryptology ePrint
  Archive}, 2016.

\bibitem{pinto2019demystifying}
S.~Pinto and N.~Santos, ``Demystifying arm trustzone: A comprehensive survey,''
  {\em ACM computing surveys (CSUR)}.

\bibitem{mcgillion2015open}
B.~McGillion, T.~Dettenborn, T.~Nyman, and N.~Asokan, ``Open-tee--an open
  virtual trusted execution environment,'' 2015.

\end{thebibliography}
